\documentclass[conference]{IEEEtran}
\IEEEoverridecommandlockouts
\usepackage[utf8]{inputenc}
\usepackage{cite}
\usepackage{overpic}
\usepackage{amsmath,amssymb,amsfonts}
\usepackage{bm}
\usepackage{algorithmic}
\usepackage{graphicx}
\usepackage{textcomp}
\usepackage{xcolor}
\usepackage{float}
\usepackage{amsthm}
\usepackage{graphicx}
\usepackage{epstopdf}
\usepackage{amsmath,bm,bbm}
\usepackage{amsfonts}
\usepackage{amssymb}
\usepackage{color}
\usepackage{multirow}
\usepackage{multicol}
\usepackage{soul,xcolor}
\usepackage{setspace}

\DeclareMathOperator{\tr}{tr}

\theoremstyle{plain}

\newcommand{\mathacr}[1]{\mathsf{#1}}

\newcommand{\vect}[1]{\mathbf{#1}}

\def\tr{\mathrm{tr}}

\def\Htran{\mbox{\tiny $\mathrm{H}$}}
\def\Ttran{\mbox{\tiny $\mathrm{T}$}}
\def\CN{\mathcal{N}_{\mathbb{C}}} 
\begin{document}
\bstctlcite{IEEEexample:BSTcontrol}
\makeatletter
\newcommand*{\rom}[1]{\expandafter\@slowromancap\romannumeral #1@}
\makeatother

\title{Cell-Free Massive MIMO in Virtualized CRAN: How to Minimize the Total Network Power?
\thanks{This work was partially funded by CELTIC-NEXT project AI4Green with the support of Vinnova, Swedish Innovation Agency.}}
\author{\IEEEauthorblockN{\"Ozlem Tu\u{g}fe Demir$^*$, Meysam Masoudi$^*$, Emil Bj\"ornson$^{*\ddagger}$, and Cicek Cavdar$^*$}
	\IEEEauthorblockA{{$^*$Department of Computer Science, KTH Royal Institute of Technology, Kista, Sweden
		} \\ {$^\ddagger$Department of Electrical Engineering, Link\"oping University, Link\"oping, Sweden
		} \\
		{Email: \{ozlemtd, masoudi, emilbjo, cavdar\}@kth.se}
	}
}
\vspace{-160pt}

\maketitle
\begin{abstract}
Previous works on cell-free massive MIMO mostly consider physical-layer and fronthaul transport aspects. How to deploy cell-free massive MIMO functionality in a practical wireless system is an open problem. This paper proposes a new cell-free architecture that can be implemented on top of a virtualized cloud radio access network (V-CRAN). 
We aim to minimize the end-to-end power consumption by jointly considering the radio, optical fronthaul, virtualized cloud processing resources, and spectral efficiency requirements of the user equipments. The considered optimization problem is cast in a mixed binary second-order cone programming form and, thus, the global optimum can be found using a branch-and-bound algorithm. The optimal power-efficient solution of our proposed cell-free system is compared with conventional small-cell implemented using V-CRAN, to determine the benefits of cell-free networking. The numerical results demonstrate that cell-free massive MIMO increases the maximum rate substantially, which can be provided with almost the same energy per bit. We show that it is more power-efficient to activate cell-free massive MIMO already at low spectral efficiencies (above 1\,bit/s/Hz).
\end{abstract}
\vspace{-2mm}
\section{Introduction}
\vspace{-2mm}

To deal with the ever-increasing mobile data traffic, one way to increase the network's traffic capacity is to enable multiple access points (APs) to serve user equipments (UEs) by coherently combining the transmit/receive signals and suppressing interference. For joint processing of UEs’ signals, it is required to centralize the digital units (DUs) that are responsible for baseband processing. Cloud radio access network (CRAN) emerged as a new architecture not only to enjoy the advanced joint transmission/reception (JT) technologies but also to exploit DU pooling in a cloud with shared housing and processing facilities. The latter leads to reduced power consumption by benefiting from the shared computational resources that can be allocated among the APs depending on their loads. This virtualization results in the virtualized CRAN (V-CRAN) architecture \cite{wang2017virtualized,wang2016joint,wang2016energy,Masoudi2020,sigwele2017energy,masoudi2019green}. 

Cell-free massive MIMO (multiple-input multiple-output) is a promising physical-layer technology which combines ultra-dense networks with JT and the low-complexity linear processing schemes from massive MIMO  \cite{cell-free-book}. By taking advantage of both joint processing and macro diversity, cell-free massive MIMO reduces the large data rate variations across the coverage area. Future networks will build on the CRAN architecture but how to implement cell-free massive MIMO on top of it remains an open research question. 

Mostly the physical-layer aspects of cell-free massive MIMO have been studied in previous work and only the radio site power consumption has been considered \cite{van2020joint}. In \cite{wang2017virtualized}, a cell-free network architecture is presented by optimizing the user-centric formation of soft base stations (BSs) defined as joint allocation of spectrum, optical wavelength, and cloud processing resources together with set of APs considering JT. End-to-end power consumption\cite{wang2017virtualized}, and network throughput\cite{wang2016joint} is optimized considering radio, optical fronthaul, and cloud processing resources. However massive MIMO is not considered in these cell-free networks. In \cite{pan2017joint}, AP selection for JT under fronthaul constraints is studied but the DU cloud power consumption is simplified as a fixed parameter. In \cite{ha2016computation}, the processing requirements in the cloud are taken into consideration, but only radio site power consumption is minimized.

  In this paper, we model the end-to-end total network power consumption for cell-free massive MIMO on V-CRAN including radio, fronthaul, and cloud resources.  We derive the DU cloud processing requirements of a cell-free massive MIMO OFDM system given required system performance. Our new network architecture will be designed to minimize the end-to-end power consumption by joint allocation of transmit powers, optical fronthaul resources, processing resources of the DU cloud, and the set of APs serving the UEs to meet their quality of service (QoS) requirements. We cast the problem in a mixed binary second-order cone programming form, which can be optimally solved. Through numerical simulations, we show when cell-free massive MIMO is more power-efficient and  provides a higher data rate than small-cell networks where each UE is served by only one AP. 
  
\vspace{-2mm}

\section{V-CRAN for Cell-free Massive MIMO}
\vspace{-2mm}

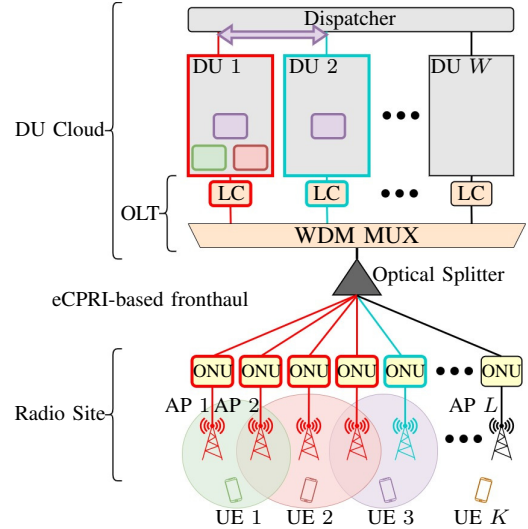
\begin{figure}[t]
\vspace{2mm}
	\hspace{1.5cm}
		\begin{overpic}[width=5.6cm,tics=10]{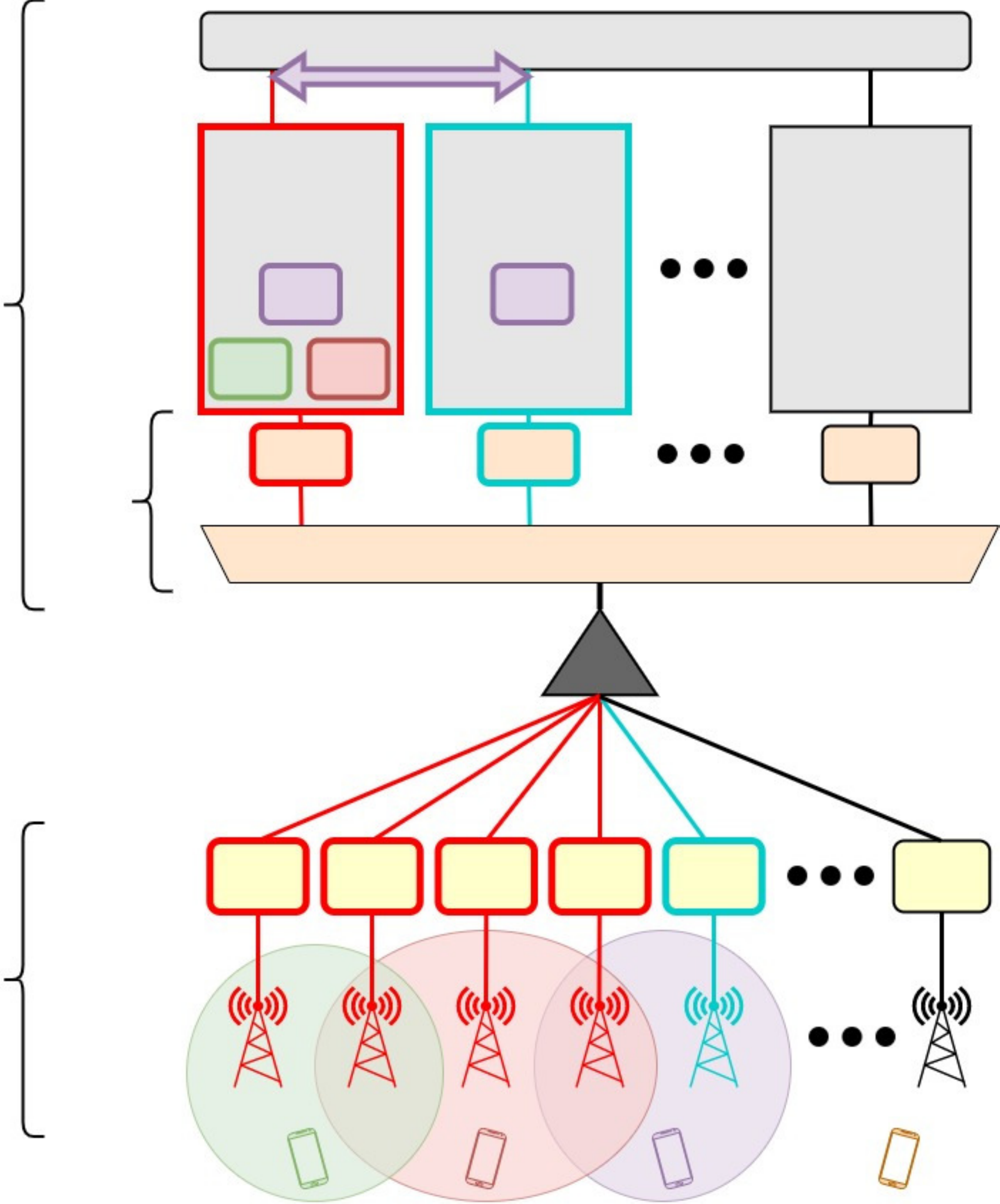}
			\put(22,-3){\footnotesize UE $1$}
				\put(36,-3){\footnotesize UE $2$}
					\put(52,-3){\footnotesize UE $3$}
						\put(69,-3){\footnotesize UE $K$}
    		\put(12,19){\footnotesize AP $1$}
			\put(21.7,19){\footnotesize AP $2$}
				\put(68,19){\footnotesize AP $L$}
		\put(17.5,86){\footnotesize DU $1$}		\put(36.5,86){\footnotesize DU $2$}
			\put(64.5,86){\footnotesize DU $W$}
	\put(17.6,26){\scriptsize ONU}
	\put(27.1,26){\scriptsize ONU}
	\put(36.6,26){\scriptsize ONU}
	\put(46,26){\scriptsize ONU}
	\put(55.5,26){\scriptsize ONU}
	\put(74.5,26){\scriptsize ONU}
	\put(22.5,61){\footnotesize LC}
	\put(41.3,61){\footnotesize LC}
	\put(69.8,61){\footnotesize LC}
\put(3.3,57){\footnotesize OLT}
\put(-17.5,73.5){\footnotesize DU Cloud}
\put(-17.6,17.5){\footnotesize Radio Site}
\put(-10.2,40){\footnotesize eCPRI-based fronthaul}
\put(53,45){\footnotesize Optical Splitter}

			\put(37.6,52.5){\small WDM MUX}
						\put(39.5,95.5){\footnotesize Dispatcher}
		\end{overpic}
	\vspace{0mm}
	\caption{Illustration of the considered V-CRAN architecture on which the cell-free massive MIMO system is implemented.}
	\label{fig:architecture}
	\vspace{-4mm}
\end{figure}

We consider a cell-free massive MIMO system that is built on top of the V-CRAN architecture shown in Fig.~\ref{fig:architecture}. There are $L$ APs and $K$ UEs that are arbitrarily distributed in the coverage area. All UEs have a single antenna while each AP is equipped with $N$ antennas. All the APs are connected to the DU cloud within a V-CRAN network architecture \cite{wang2017virtualized}, via fronthaul connections. A set of APs serves each UE to satisfy the corresponding spectral efficiency (SE) requirement. Let $x_{k,l}\in\{0,1\}$ be the binary variable denoting whether UE $k$ is served by AP $l$ or not. For example, in Fig.~\ref{fig:architecture}, the colored circular regions for each UE indicate the APs that are serving them. In this paper, we will optimize these subsets to satisfy the QoS requirement of each UE in an aim to minimize the total network power consumption.

We consider the recently released evolved CPRI (eCPRI) specification for the fronthaul transmission with the physical layer (PHY)-radio frequency (RF) functional split E  \cite{perez2018fronthaul} to fully benefit from the power-saving feature of CRAN. According to this split, the APs only perform RF processing and transmit (receive) the pure sampled and quantized baseband signals to (from) the DU cloud via fronthaul links. All the remaining processing is done in the DU cloud.  
 
A time- and wavelength-division multiplexed passive optical network (TWDM-PON) is utilized as the fronthaul transport network to carry eCPRI packets to meet the high capacity requirements of the fronthaul transmission in a cell-free network \cite{wang2017virtualized,wang2016energy}. As shown in Fig.~\ref{fig:architecture}, each AP is connected to one optical network unit (ONU) that is assigned to one of the multiple wavelengths in fiber communications. Each wavelength can be shared by more than one AP using time-division multiplexing. In the DU cloud, there exists an optical line terminal (OLT) with a WDM multiplexer (WDM MUX) and multiple line cards (LCs), each of which is connected to one DU. Each LC serves only one wavelength and, thus, each AP's signals are received at or transmitted from the DU that uses the same wavelength. In Fig.~\ref{fig:architecture}, the same colors are used to show which APs are connected to which DU. For example, DU 1 is responsible for the first four APs' data whereas AP 5 is connected to DU 2. 
    
In the DU cloud, there are $W$ stacks of DUs. We assume the general purpose processors (GPPs) are used for baseband processing due to their processing capabilities and programmability, which allows virtualization. The workload of each DU is routed via a dispatcher that is controlled by a global cloud controller \cite{sigwele2017energy}. In Fig.~\ref{fig:architecture}, UE 1 is served by AP 1 and AP 2, which are all connected to DU 1. In this case, UE 1 can be served using potentially only DU 1 without any need for data exchange among DUs. However, since the APs that serve UE 3 are connected to different DUs, the connection should be activated between DU 1 and DU 2 for the sharing of UE 3 data and control signal.

Shutting down the unused active DUs together with the corresponding LCs, we can save power by minimizing the active idle power of the network \cite{sigwele2017energy}. However, there is a trade-off between minimizing the number of active DUs and satisfying the required SE performance. 

\vspace{-2mm}

\section{Cell-Free Massive MIMO OFDM}
\vspace{-2mm}

A cell-free massive MIMO system with time-division duplex and frequency-selective channels is considered. 
OFDM is utilized where the sampling rate is $f_s$ and the total bandwidth is $B$. The total number of subcarriers is $N_{\rm DFT}$, which is also the dimension of the discrete Fourier transform (DFT), while the number of used subcarriers is $N_{\rm used}\leq N_{\rm DFT}$. Each OFDM symbol has a duration of  $T_{ s}$ seconds. We assume block fading channel model. The channels are constant and take an independent realization in each coherence block that consists of $N_{\rm slot}$ OFDM symbols. We denote the number of consecutive subcarriers in each coherence block by $N_{\rm smooth}$ so that the channels in each coherence block have an approximately constant (smooth) frequency response. The number of useful samples (channel uses) in each coherence block is $\tau_c=N_{\rm smooth}N_{\rm slot}$. 

We consider distributed downlink operation where the channel estimates corresponding to each AP are used for distributed per-AP precoding \cite[Sec. 6.2]{cell-free-book}. In this way, we can divide the UE-related processing tasks into small independent blocks that can be virtualized in the DU cloud to minimize the number of active DUs.
\footnote{This is not possible in the centralized cell-free operation where a higher-dimensional precoding should be computed jointly for all the serving APs, based on pooling of channel estimates, and then applied to a particular UE in the same DU.} Note that the fronthaul requirements scale with the number of AP antennas and do not depend on the number of UEs that an AP can serve.

Since we focus on the downlink operation,  each coherence block is divided into two phases: i) uplink training with $\tau_p$ samples  and ii) downlink payload data transmission with $\tau_d=\tau_c-\tau_p$ samples. All the UEs are served on the same time-frequency resources using spatial multiplexing. To make the notation simpler, we will focus on an arbitrary time-frequency resource block as in \cite{cell-free-book}.
 \vspace{-2mm}

\subsection{Channel Model}
\vspace{-1mm}

We let $\vect{h}_{kl}\in \mathbb{C}^{N}$ denote the channel from UE $k$ to AP $l$ in an arbitrary coherence block. The channels are modeled using correlated Rayleigh fading, i.e.,  $\vect{h}_{kl}\sim\mathcal{N}_{\mathbb{C}}({\bf 0}_N,{\bf R}_{kl})$ and they are independent for different UEs and APs. The correlation matrix ${\bf{R}}_{kl}\in \mathbb{C}^{N \times N}$ is determined by the spatial correlation of the channel $\vect{h}_{kl}$ between the antennas of AP $l$ and the corresponding average channel gain, which is denoted by $\beta_{kl}=\tr({\bf R}_{kl})/N$. The channel gain is dependent on the large-scale effects such as geometric attenuation and shadowing. 

To perform coherent transmit processing, the channels need to be estimated in each coherence block.   
The pilot sequences are assigned to the UEs in a deterministic way and the UEs simultaneously send their pilots to the APs. After initial RF processing at the radio site, the received signals are transmitted to the DU cloud and the channel estimation is performed there for each coherence block. Using all the received pilot signals, the minimum mean-squared error (MMSE) channel estimate $\widehat{\vect{h}}_{kl}$ of the channel $\vect{h}_{kl}$ can be computed using \cite[Corol.~4.1]{cell-free-book}.

\vspace{-1mm}

 \subsection{Downlink Data Transmission}
 \vspace{-1mm}

Let $\varsigma_i\in \mathbb{C}$ denote the downlink data signal of UE $i$ with $\mathbb{E} \{|\varsigma_i|^2 \} = 1$. Let the normalized (in terms of average power) precoding vector and transmit power corresponding to UE $i$ and AP $l$ for $x_{i,l}=1$ be $\vect{w}_{il}\in \mathbb{C}^N$ and $p_{il}\geq 0$, respectively.  In the DU cloud, the frequency-domain precoded signal to be transmitted from AP $l$ is constructed as
\vspace{-1.5mm}
\begin{equation}
    \vect{x}_l=\sum_{i=1}^K\sqrt{p_{il}}x_{i,l}\vect{w}_{il}\varsigma_i\in \mathbb{C}^{N}.
    \vspace{-1.5mm}
\end{equation}
  
The received frequency-domain downlink signal at UE $k$ is
  \vspace{-1.5mm}
\begin{equation}  \label{eq:downlink-received-UEk}
y_k^{\rm{dl}} = \sum_{l=1}^{L}  {\vect{h}}_{kl}^{\Ttran} \vect{x}_l + n_k =\sum_{l=1}^{L} \sum_{i=1}^{K}\sqrt{p_{il}}  x_{i,l} {\vect{h}}_{kl}^{\Ttran} \vect{w}_{il} \varsigma_i + n_k
  \vspace{-1.5mm}
\end{equation}
where $n_k \sim \CN(0,\sigma^2)$ is the receiver noise. The downlink achievable SE of UE $k$ at each resource block can be computed using \cite[Corr.~6.3 and Sec. 7.1.2]{cell-free-book} as
  \vspace{-1.5mm}
	\begin{equation} \label{eq:downlink-rate-expression-level2}
	\begin{split}
	\mathacr{SE}_{k} = \frac{\tau_d}{\tau_c} \log_2  \left( 1 + \mathacr{SINR}_{k}  \right) \quad \textrm{bit/s/Hz}
	\end{split}
	  \vspace{-1.5mm}
	\end{equation}
	with the effective signal-to-interference-plus-noise ratio (SINR) given by
	  \vspace{-1.5mm}
\begin{equation}\label{eq:SINR-downlink-2}
\mathacr{SINR}_{k}=\frac{\left|\vect{b}_k^{\Ttran}{\boldsymbol{\rho}}_k\right|^2}{\sum\limits_{i=1}^K{\boldsymbol{\rho}}_i^{\Ttran}{\vect{C}}_{ki}{\boldsymbol{\rho}}_i+\sigma^2}
  \vspace{-2.5mm}
\end{equation}
where
  \vspace{-2.5mm}
\begin{align}
 &{\boldsymbol{\rho}}_k=\left [ \sqrt{p_{k1}}x_{k,1} \, \ldots \, \sqrt{p_{kL}} x_{k,L} \right ]^{\Ttran} \in \mathbb{R}_{\geq 0}^L,  \\
  &{\vect{b}}_k \in \mathbb{R}_{\geq 0}^L, \quad
    \left[ {\vect{b}}_{k} \right]_{l}= \mathbb{E} \left\{ \vect{h}_{kl}^{\Ttran}{\vect{w}}_{kl}\right\},{\vect{C}}_{ki}\in \mathbb{C}^{L \times L},  \\ 
 & \left[ {\vect{C}}_{ki} \right]_{lr}\!=\!\begin{cases}
  \mathbb{E} \left\{ \vect{h}_{kl}^{\Ttran}{\vect{w}}_{kl}{\vect{w}}_{kr}^{\Htran}\vect{h}_{kr}^*\right\}\!-\! \left[ {\vect{b}}_{k} \right]_{l} \left[ {\vect{b}}_{k} \right]_{r}^*, & i=k,  \\
  \mathbb{E} \left\{ \vect{h}_{kl}^{\Ttran}{\vect{w}}_{il}{\vect{w}}_{ir}^{\Htran}\vect{h}_{kr}^*\right\}, & i\neq k. \end{cases} 
\end{align}

In this paper, we will use local partial MMSE (LP-MMSE precoding) \cite[Sec. 6.2.2]{cell-free-book}, which suppresses interference to the UEs with the strongest channel per pilot at each AP using the channel estimates $\{\widehat{\vect{h}}_{il} \}$.

  \vspace{-1mm}
\section{Power Consumption Modeling}
  \vspace{-1.5mm}
For the considered V-CRAN architecture, we do not stick to specific hardware but instead adopt a generic power model that can be applied to the different technologies.
The network power consumption consists of two main components: i) the radio site power consumption that includes the AP power consumption $P_{{\rm AP},l}$, for $l=1,\ldots,L$ and the ONU power consumption, $P_{\rm ONU}$; and ii) the DU cloud power consumption, $P_{\rm DU-C}$ \cite{Masoudi2020}. The total power consumption is given as

\begin{equation}
    P_{\rm tot} = \sum_{l=1}^L P_{{\rm AP},l} + \sum_{l=1}^Lz_lP_{\rm ONU} + P_{\rm DU-C} \label{eq:total-power-consumption}
      \vspace{-2mm}
\end{equation}
where the binary variable $z_{l}$ indicates whether AP $l$ is active or not. If it is active, then $z_l=1$ and otherwise $z_l=0$. The AP power consumption for a particular AP $l$, i.e., $P_{{\rm AP},l}$ is given by \cite{Auer2011,Masoudi2020}
\vspace{-2mm}
\begin{align} \label{eq:AP-l-energy}
    &P_{{\rm AP},l} =\\& z_l \left(P_{{\rm AP},0}+\Delta^{\rm tr}\sum_{k=1}^Kx_{k,l}p_{kl}\right) 
     \stackrel{(a)}=  z_l P_{{\rm AP},0}+\Delta^{\rm tr}\sum_{k=1}^Kx_{k,l}p_{kl}\nonumber
\end{align}
where $P_{{\rm AP},0}$ is the static power consumption of each AP when there is no transmission at the active mode and $\sum_{k=1}^Kx_{k,l}p_{kl}$ is the transmit power of AP $l$. The load-dependent power consumption is modeled by the slope $\Delta^{\rm tr}$. In (a), we have simplifed the equation by noting that when $z_l$ is zero, $x_{k,l}$, $\forall k$ become zero by definition. It is worth mentioning that if an AP is not active, i.e., $z_l=0$, we turn it off to save power.  

The total power consumption at the DU cloud is computed  using the load-dependent GPP power consumption model from \cite{sigwele2017energy} as
  \vspace{-3mm}
\begin{align} \label{eq:CPU-power-consumption}
    P_{\rm DU-C} =& \frac{1}{\sigma_{\rm cool}}\Bigg( P_{\rm disp}+ P_{\rm OLT}\sum_{w=1}^W w\ell_{w}\nonumber\\
    &+\sum_{w=1}^Wwd_wP_{\rm DU-C,0}^{\rm proc}+\Delta^{\rm proc}_{\rm DU-C} \frac{C_{{\rm DU-C}}}{C_{\rm  max}}\Bigg),
\end{align}
where $0<\sigma_{\rm cool}\leq1$ is the cooling efficiency of the DU cloud.\footnote{ Cooling power is modeled as a load-dependent parameter since we focus on a small-scale network with a nano-scaled cloud that is potentially co-located with a macro cell BS \cite{Auer2011}.} $P_{\rm disp}$ and $P_{\rm OLT}$ are the power consumptions of the GPP dispatcher and each OLT module per DU \cite{Masoudi2020}, respectively.  The binary variable $\ell_{w}\in \{0,1\}$ is one if the LCs of $w$ number of DUs are active and zero otherwise. Similarly, $d_w\in \{0,1\}$ is one if  $w$ number of DUs are active for either processing or fronthaul connection to the APs, and zero otherwise. This particular definition of the binary variables is to ensure that the constraints can be written as linear functions of the binary variables in the optimization problem we will consider in the next section. It is worth mentioning that the LC of an active DU may be inactive if the corresponding DU participates only in the processing that is redirected to it coming from other DUs.  $P_{\rm DU-C,0}^{\rm proc}$ is the idle mode processing power consumption of each active DU at the cloud corresponding to zero utilization. $\Delta^{\rm proc}_{\rm DU-C}$ is the slope of the load-dependent processing power consumption of each DU at the cloud, respectively. For each DU, the maximum processing capacity is given by $C_{\rm max}$ in giga-operations per second (GOPS) depending on the used technology. The total DU utilization is given by $0\leq C_{{\rm DU-C}} \leq W C_{\rm  max}$ in GOPS. In the following part, we will compute the required GOPS at the DU cloud for cell-free massive MIMO processing.

\subsection{GOPS Analysis of Digital Operations at the DU Cloud }

In this section, we will analyze the GOPS for digital signal processing operations of a cell-free massive MIMO system.  In the uplink, after RF processing at the APs, the quantized baseband signals are directly sent to the cloud. Then, at the DUs, cyclic prefix (CP) removal and $N_{\rm DFT}$-point DFT are performed to obtain frequency-domain signals. After resource element (RE) demapping, the remaining PHY and higher-layer operations are implemented for a particular UE. Similarly, in the downlink, after the higher-layer functions, the precoded signals are obtained. In the sequel, RE mapping, inverse DFT, and CP insertion are realized. Then, the time-domain signals are sent to the APs and RF transmission takes place.

To compute the total GOPS in the cloud, $C_{{\rm DU-C}}$, we will mainly use the results from cellular massive MIMO \cite{malkowsky2017world,desset2016massive}. In \cite{desset2016massive}, a factor two of overhead is taken in arithmetic operation calculations to account for memory operations. In the following GOPS calculations, we will also consider this by including a multiplication by two in each arithmetic operation. 
The first operation after RF processing is baseband filtering. Considering 10 taps with a polyphase filtering implementation, the corresponding complexity per AP is given in GOPS as $ C_{\rm filter} = 40Nf_s/10^9$ \cite{desset2016massive}.
The next operation is
 DFT in the uplink and inverse DFT in the downlink, which has the complexity with fast Fourier transform (FFT) as $C_{\rm DFT} =8N N_{\rm DFT}\log_2\left(N_{\rm DFT}\right)/\left(T_{ s}10^9\right)$,
which is obtained by dividing the number of required real operations by the OFDM symbol duration $T_{ s}$ \cite{malkowsky2017world}.

The GOPS of the sample-based arithmetic operations such as precoding  scale with $N_{\rm used}/T_{ s}$ \cite{malkowsky2017world}. An additional multiplying factor $\tau_d/\tau_c$ should be taken into account in precoding the downlink data since $\tau_d$ samples are precoded in each coherence block of length $\tau_c$. For the channel estimation, reciprocity calibration, and precoding computation, it scales with $N_{\rm used}/(T_{ s}\tau_c)$ since the corresponding operations are common for each sample in a coherence block of length $\tau_c$. To this end, from \cite[Sec.~6.2.2]{cell-free-book}, for LP-MMSE transmit precoding together with the required channel estimation of the served UEs by AP $l$ (and of the strongest UE per pilot), the GOPS (in terms of real multiplications) is computed as
\vspace{-1mm}
\begin{align} \label{eq:complexity-PMMSE}
    C_{{\rm prec},l} = &\underbrace{\frac{N_{{\rm used}}}{T_{ s}\tau_c10^9}\left(8N\tau_p^2+8N^2\left(\tau_p+\sum_{i=1}^Kx_{i,l}\right)\right)}_{\textrm{Channel estimation}}  \nonumber\\&\hspace{-16mm} +\! \underbrace{\frac{N_{{\rm used}}\tau_d}{T_{ s}\tau_c10^9}8N\sum_{i=1}^Kx_{i,l}}_{\textrm{Precoding}}+  \underbrace{\frac{N_{{\rm used}}}{T_{ s}\tau_c10^9}8N\sum_{i=1}^Kx_{i,l}}_{\textrm{Reciprocity calibration}}\nonumber\\
    &\hspace{-16mm}+\!\underbrace{\frac{N_{{\rm used}}}{T_{ s}\tau_c10^9}\!\left(\!\left(4N^2\!+\!4N\right)\tau_p\!+\!8N^2\sum_{i=1}^Kx_{i,l}\!+\!\frac{8\!\left(\!N^3\!-\!N\!\right)}{3}\right)\!}_{\textrm{Precoding computation}} \end{align}
where we have also included the complexity of applying precoding and reciprocity calibration from \cite{malkowsky2017world}. It is worth mentioning that local precoding computation $C_{{\rm prec},l}$ in \eqref{eq:complexity-PMMSE} can be implemented at any other DU $w$ different than the connected DU to AP $l$  (benefiting from DU cloud sharing and virtualization via GPP dispatcher).

In addition to precoding, there are other GOPS regarding OFDM modulation/demodulation, mapping/demapping, channel coding, higher-layer control and network operations. These can be computed using the flexible power modeling in \cite{Debaillie2015a}. Let $C_{\rm other,AP}$ and $C_{\rm other,UE}$ denote the GOPS for the other operations which scale with the number of active APs and the number of UEs that are served by each AP, respectively. There is also a fixed GOPS for UE operations, which are independent of the number of serving APs. This is is represented by $\mathcal{F}$. Then, the total GOPS in the cloud is given by
\begin{align}
    C_{\rm DU-C}=& \sum_{l=1}^Lz_l\left(C_{\rm filter}+C_{\rm DFT}+ C_{{\rm prec},l}+C_{\rm other,AP}\right) \nonumber\\
    &\hspace{-19mm}+C_{\rm other,UE}\!\!\sum_{l=1}^L\!\sum_{k=1}^K\!x_{k,l}\!+\!\mathcal{F}\! 
    \triangleq\!\mathcal{Z}\! \sum_{l=1}^L\! z_l\!  +\! \mathcal{X}\! \sum_{l=1}^L\!\sum_{k=1}^K\!x_{k,l}\!+\!\mathcal{F}\!\!
\end{align}
where the constant parameters $\mathcal{Z}$ and $\mathcal{X}$ are defined for ease of notation in the optimization problem.

\vspace{-1mm}

\section{Power-Efficient AP Selection, DU and Power Allocation}
\vspace{-1mm}
In this section, we will introduce the proposed optimization problem that minimizes the total power consumption. The aim is to decide which APs serve which UEs, i.e., the binary variables $x_{k,l}\in \{0,1\}$, the transmit powers allocated to the UEs, i.e., $p_{kl}$, which APs are active and connected to the cloud, i.e., $z_{l}\in \{0,1\}$, and the number of active LCs and DUs, i.e., $\ell_w,d_w\in\{0,1\}$ in the cloud. 
We note that the considered optimization problem is a mixed integer program since the AP selection together with power allocation and DU allocation  is considered, which has a combinatorial nature. To express both the objective function and the constraints in a mixed binary linear or conic form, we introduce the following additional optimization variables:
\begin{itemize}
    \item  ${\boldsymbol{\rho}}_k=\left [ \sqrt{p_{k1}}x_{k,1} \, \ldots \, \sqrt{p_{kL}} x_{k,L} \right ]^{\Ttran} = \left [\rho_{k,1} \, \ldots \, \rho_{k,L}\right]^{\Ttran} $
  \item  ${\boldsymbol{\rho}}^{\prime}_l=\left [ \sqrt{p_{1l}}x_{1,l} \, \ldots \, \sqrt{p_{Kl}} x_{K,l} \right ]^{\Ttran} = \left [\rho_{1,l} \, \ldots \, \rho_{K,l}\right]^{\Ttran} $
 \end{itemize}

With PHY-RF functional split E, the required fronthaul data rate for each AP is given as $ R_{\rm fronthaul} = 2f_sN_{\rm bits}N$, where $N_{\rm bits}$ is the number of bits to quantize the signal samples \cite{perez2018fronthaul}. Due to limited capacity of each wavelength in TWDM-PON, which is denoted by $R_{\rm max}$, we can assign at most $W_{\rm max}\triangleq\lfloor R_{\rm max}/  R_{\rm fronthaul}\rfloor$ APs to each wavelength and, hence, to each DU $w$, for $w=1,\ldots,W$.

In the considered network power consumption minimization problem, we assume each UE $k$ has a SE request with the corresponding minimum SINR requirement $\gamma_k$. Hence, we have QoS constraints in the form of $\mathacr{SINR}_{k}\geq \gamma_k$ for each UE $k$. The optimization problem can be cast using the introduced optimization variables as
\begin{subequations} \label{eq:optimization1}
\begin{align} 
&\underset{ z_l, \ell_w, d_w, x_{k,l} \in \{0,1\} }{\textrm{minimize}} \quad  \frac{P_{\rm disp}}{\sigma_{\rm cool}} +  \sum_{l=1}^Lz_{l}P_{l}+\Delta^{\rm tr}\sum_{l=1}^L\sum_{k=1}^K\rho_{k,l}^2\nonumber\\&+ \frac{P_{\rm OLT}}{\sigma_{\rm cool}}\sum_{w=1}^W w\ell_{w}+\frac{P_{\rm DU-C,0}^{\rm proc}}{\sigma_{\rm cool}}\sum_{w=1}^Wwd_w\nonumber\\
&+\frac{\Delta^{\rm proc}_{\rm DU-C}\mathcal{X}}{\sigma_{\rm cool}C_{\rm max}}\sum_{l=1}^L\sum_{k=1}^K x_{k,l}+\frac{\Delta^{\rm proc}_{\rm DU-C}\mathcal{F}}{\sigma_{\rm cool}C_{\rm max}} \label{eq:Xobjective} 
\end{align}
\begin{align}
\textrm{subject to} \nonumber \\
&\hspace{-10mm}\frac{\left|\vect{b}_k^{\Ttran}{\boldsymbol{\rho}}_k\right|^2}{\sum\limits_{i=1}^K{\boldsymbol{\rho}}_i^{\Ttran}{\vect{C}}_{ki}{\boldsymbol{\rho}}_i+\sigma^2}\geq \gamma_k, \quad \forall k, \label{eq:Xconstraint1} \\
& \hspace{-10mm} \sum_{l=1}^Lz_{l}\leq W_{\rm max}W, \label{eq:Xconstraint2} \\
& \hspace{-10mm} \frac{\sum_{k=1}^Kx_{k,l}}{K} \leq z_{l} \leq \sum_{k=1}^Kx_{k,l},  \quad \forall l, \label{eq:Xconstraint4} \\
& \hspace{-10mm} \sum_{w=1}^{W}(w-1)\ell_{w}\leq \frac{\sum_{l=1}^Lz_l}{W_{\rm max}} \leq  \sum_{w=1}^{W}w\ell_w, \label{eq:Xconstraint5} \\
& \hspace{-10mm} \mathcal{Z} \sum_{l=1}^Lz_{l}  +\mathcal{X}\sum_{l=1}^L\sum_{k=1}^K x_{k,l}+\mathcal{F}\leq C_{\rm  max} \sum_{w=1}^W wd_w, \label{eq:Xconstraint6} \\
&  \hspace{-10mm}\sum_{w=1}^W \ell_w= 1, \quad \sum_{w=1}^W d_{w} = 1, \label{eq:Xconstraint8} \\
& \hspace{-10mm} \sum_{w=1}^W w\ell_w \leq \sum_{w=1}^W wd_w,  \label{eq:Xconstraint9}\\
&\hspace{-10mm} 0\leq\rho_{k,l}\leq \sqrt{p_{\rm max}}x_{k,l}, \quad \forall k, \forall l, \label{eq:Xconstraint10} \\
& \hspace{-10mm}\left\Vert {\boldsymbol{\rho}}^{\prime}_l \right\Vert\leq \sqrt{p_{\rm max}}z_l, \quad  \forall l, \label{eq:Xconstraint11} 
\end{align}
\end{subequations}
with $P_l=  P_{{\rm AP},0} +P_{\rm ONU}+\Delta^{\rm proc}_{\rm DU-C}\mathcal{Z}/ \left(C_{\rm  max}\sigma_{\rm cool}\right)$.
The constraints in \eqref{eq:Xconstraint1} are to guarantee that each UE's minimum SINR requirement is satisfied. The constraint in \eqref{eq:Xconstraint2} guarantees that the number of active APs does not exceed the maximum allowable number determined by the fronthaul limitations. The constraints in \eqref{eq:Xconstraint4} relate the binary variables $x_{k,l}$ and $z_l$, i.e., an AP is active if and only if it serves at least one UE. The constraint in \eqref{eq:Xconstraint5} connects the number of active APs to the the number of required active LCs. The constraint in \eqref{eq:Xconstraint6} guarantees that the total GOPS does not exceed the processing capability of active DUs. The constraints in \eqref{eq:Xconstraint8} are to satisfy that $\ell_w$ and $d_w$ are only one for one value of $w$ since these binary variables are one when the number of active LCs or DUs is equal to their sub-index. The constraint in \eqref{eq:Xconstraint9} is to ensure that the number of active DUs is always greater than or equal to the number of active LCs. The constraints in \eqref{eq:Xconstraint10} guarantee that the square root of the power coefficient for UE $k$ and AP $l$ is zero if UE $k$ is not served by AP $l$. Here, $p_{\rm max}$ is the maximum transmit power budget of each AP and when $x_{k,l}=1$, this constraint does not limit $\rho_{k,l}$. \eqref{eq:Xconstraint11} represents the per-AP transmit power constraints. Note that the SINR constraints in \eqref{eq:Xconstraint1} can be re-written in second-order cone form \cite[Sec.~7.1.2]{cell-free-book}. As a result, the overall optimization problem is a mixed binary second-order cone programming problem, which has a convex structure except for the binary constraints. Hence, the global optimum solution is obtained by the branch-and-bound algorithm \cite{gurobi}. It is known that the complexity grows exponentially with the number of discrete variables, which in our case means the number of APs, UEs, and DUs.  Therefore, we limit ourselves with a small-scale simulation setup to find the optimal solution.

  \vspace{-4mm}
\section{Numerical Results and Discussion}
  \vspace{-1.5mm}
In this part, we quantify the performance gain of the proposed cell-free architecture, implemented on top of V-CRAN, compared to a conventional small-cell system where each UE is only served by one AP. To obtain a fair comparison, we consider the same V-CRAN model and obtain the power-optimal DU and power allocation solution by solving \eqref{eq:optimization1} for small-cell system by adding an additional constraint $\sum_{l=1}^L x_{k,l}=1$, $\forall k$ to guarantee that only one AP is transmitting data to each UE. Hence, the problem for small-cell is more restrictive than its cell-free counterpart. 
The simulation parameters are outlined in Table~\ref{tab:simulation} and they are mainly set from the works \cite{malkowsky2017world,Auer2011,wang2016energy,Masoudi2020,sigwele2017energy,simeonidou2020dynamic}. In particular, we consider pico-cell power parameters from \cite{Auer2011}. The GOPS/Watt for each of $W=4$ DUs in the cloud is 2.434 according to 2x Intel Xeon E5-2683 v4 processor from \cite[Tab.~1]{simeonidou2020dynamic}. Since we consider a relatively small-scale setup, the idle power $P_{\rm DU-C,0}^{\rm proc}$ and the slope $\Delta^{\rm proc}_{\rm DU-C}$ are scaled linearly such that each DU has $C_{ \rm max}=180$\,GOPS as in \cite{sigwele2017energy}. The deployment and radio site parameters are as in the running example of \cite[Sec.~5.3]{cell-free-book} except for the parameters that are listed in Table~\ref{tab:simulation}.

\begin{table}[t]
	\footnotesize
		\vspace{-0.3cm}
	\caption{Simulation Parameters.}  \label{tab:simulation}
	\vspace{-0.3cm}
	\centering
	\begin{tabular}{|c|c|c|c|}
		\hline
	 $L$, $K$, $N$, $W$&   16, 8, 4, 4    &    
	 $f_s$ &  30.72\,MHz    \\ \hline
	 $B$  & 20\,MHz    &
	 $N_{\rm DFT}$, $N_{\rm used}$ & 2048, 1200 \\   \hline
	 $T_s$ & 71.4\,$\mu$s &
	 $N_{\rm smooth}$, $N_{\rm slot}$ & 12, 16 \\ \hline
	 $\tau_c$, $\tau_p$ & 192, 8 &
Size of coverage area &  1\,km $\times$\,1\,km      \\ 
	 \hline 
	 $P_{{\rm AP},0}$, \ $\Delta^{\rm tr}$  &  6.8$N$\,W, \ 4  &    
	  Pilot power, \ $p_{\rm max}$ & 100\,mW, \   1\,W    \\ 
	 \hline  
 $P_{\rm disp}$, \ $\sigma_{\rm cool}$  & 120\,W, \ 0.9 &
  $P_{\rm ONU}$, \ $P_{\rm OLT}$ & 7.7\,W, \ 20\,W  \\  \hline
 $P_{\rm DU-C,0}^{\rm proc}$ & 20.8\,W &
 $\Delta^{\rm proc}_{\rm DU-C}$ & 74\,W \\ \hline
 $C_{ \rm max}$ & 180\,GOPS &
 $R_{\rm max}$, $N_{\rm bits}$ & 10\,Gbps, 12 \\ \hline
	\end{tabular}
\vspace{-5mm}
\end{table}

\begin{figure}[t]
	\includegraphics[trim={6mm 2mm 5mm 8mm},clip,width=3.3in]{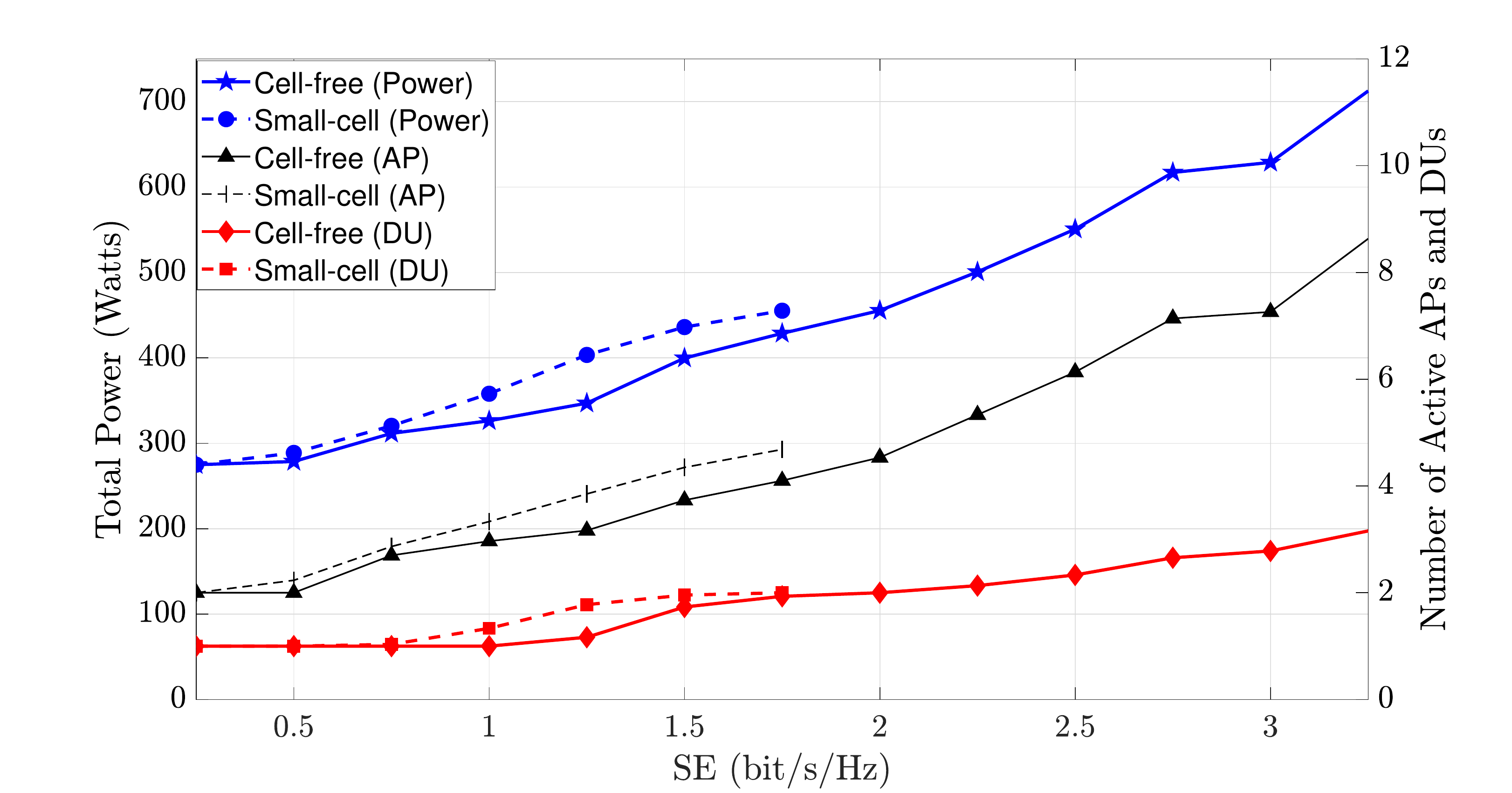}
	\vspace{-0.4cm}
	\caption{The total power (on the left y-axis) and the number of active APs and DUs (on the right y-axis) versus the SE requirement per UE.}
	\label{fig:fig1}
	\vspace{-2.5mm}
\end{figure}

We consider 30 random AP and UE locations and Fig.~\ref{fig:fig1} shows the average total power consumption and the number of active APs and DUs for the optimal power-efficient solutions for a given SE requirement that is assumed to be the same for every UE. For both cell-free and small-cell network, we consider the same random setups. However, due to the SINR constraints in \eqref{eq:Xconstraint1}, the optimization problem is not guaranteed to be feasible. In the figure, the average is taken out of all feasible setups at each point and it is only plotted when the feasibility ratio is greater than 50\%. 
One may think that the considered SE values are relatively small for a single UE alone. The reason is that all $K=8$ UEs (including the UEs with the most unfortunate channel conditions) achieve the same particular SE, which limits the maximum feasible SE. As the plot shows, when the SE requirement is greater than 1.75, the small-cell system cannot guarantee reliable performance due to infeasibility. On the other hand, the cell-free system benefits from user-centric JT to support the UEs with much higher SEs. For small SE values, the power consumption is almost the same for both systems. The reason is that for some setups, the cell-free optimization problem results in small-cell solution making their power consumption the same. However, as the SE increases, more APs and DUs are activated to serve UEs when using the small-cell system. This results in increased power consumption compared to cell-free massive MIMO as shown in the figure for the SE range $[0.5,1.75]$. The maximum power saving is $14\%$ when the SE requirement is 1.25 bit/s/Hz. In conclusion, cell-free massive MIMO results in less or equal power consumption as for a small-cell system to guarantee a certain SE requirement. In Table~\ref{tab:simulation2}, we present the average of the maximum rate that can be provided to all the UEs and the corresponding energy per bit values for cell-free and small-cell systems. Cell-free massive MIMO provides around 1.7 times more rate to the UEs with almost the same energy per bit in comparison to small-cell system. 

\begin{table}[t]
	\footnotesize
		\vspace{-0.3cm}
	\caption{Maximum Provided Rate and Energy Per Bit.}  \label{tab:simulation2}
	\vspace{-0.3cm}
	\centering
	\begin{tabular}{|c|c|c|}
		\hline
	&   Maximum Rate (Mbps)   &  Energy Per Bit (Joule/bit) 
	   \\ \hline
	Cell-free &  440 &   1.9$\cdot 10^{-6}$  \\ \hline 
	Small-cell & 260 &  2$\cdot 10^{-6}$ \\ \hline 
	\end{tabular}
\vspace{-6.5mm}
\end{table}

\begin{figure}[t]
	\includegraphics[trim={6mm 2mm 5mm 8mm},clip,width=3.3in]{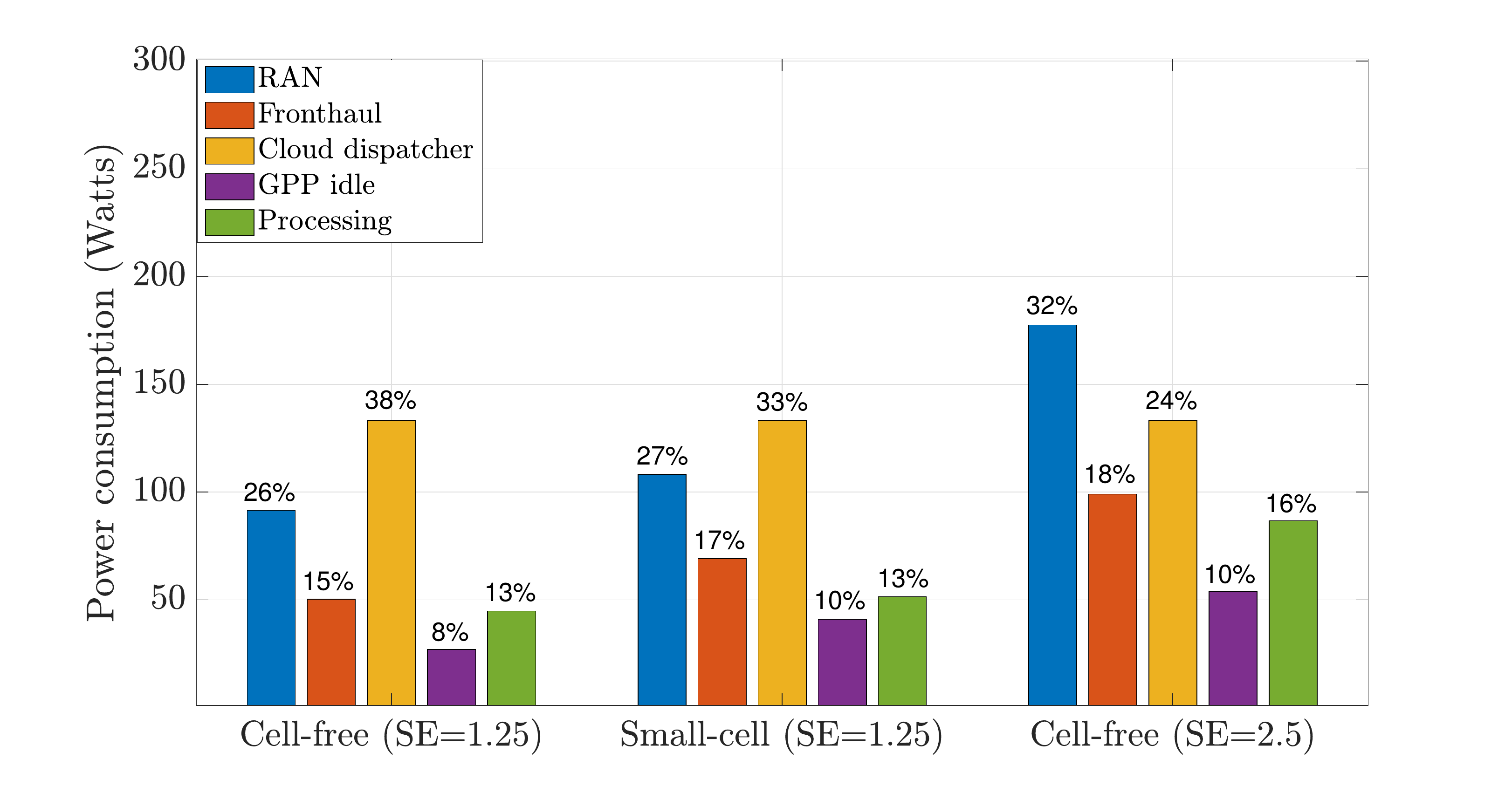}
	\vspace{-0.5cm}
	\caption{Power consumption breakdown for cell-free (two different SE requirements) and small-cell systems.}
	\label{fig:fig2}
	\vspace{-0.65cm}
\end{figure}

In Fig.~\ref{fig:fig2}, we show the average power consumption breakdown for the small-cell and cell-free systems for different SE requirements. For the particular SE requirement of 1.25 bit/s/Hz, cell-free massive MIMO provides reduced power consumption.    
This is due to reduced RAN, fronthaul, and cloud processing power consumption when cell-free operation is used. For this SE requirement, the system is lightly loaded so that the static cloud dispatcher power dominates the total power consumption. On the other hand, when the SE is increased to 2.5 bit/s/Hz (where we do not include results of small-cell network due to the corresponding infeasible  problem), the power consumption of each network component increases while the dominating component is RAN. This is a natural consequence of more activated APs to satisfy the SE requirement of the UEs.

  \vspace{-1.5mm}
\section{Conclusions}
  \vspace{-1.5mm}
In this paper, we have derived the end-to-end network power consumption of our proposed cell-free massive MIMO architecture that can be implemented on top of V-CRAN. We have obtained the optimal power-efficient AP/DU selection and transmit power coefficients under fronthaul transport limitations and SE requests. 
The proposed cell-free system is advantageous over conventional small cells for both maximum rate and minimum power consumption (up to $14\%$ saving). On the other hand, for very small SE requests, the performance of cell-free and small-cell systems are the same, since the cell-free functionality is not activated. As an initial step in this paper, we have considered a small-scale setup to show the benefits of joint and coherent cell-free transmission. Even in a small setup like that, cell-free massive MIMO increases the maximum provided rate by 1.7 with less energy per bit in comparison to small cells. When the required SE is high (more than 1.75 bit/s/Hz), the small-cell scenario may not be feasible while by activating cell-free massive MIMO we can obtain a feasible and power-efficient solution. As a future work, our aim is to devise a low-complexity meta-heuristic algorithm to see the benefits of cell-free system in a larger setup and extend our framework to a dynamic environment.

\vspace{-12pt}
\bibliographystyle{IEEEtran}
\bibliography{IEEEabrv,refs}

\end{document}